\documentclass[reprint,prb,showpacs,preprintnumbers,amsmath,amssymb,floatfix,twocolumns,footinbib]{revtex4-1}
\newcommand{\ket}[1]{\ensuremath{\left|{#1}\right\rangle}}

\usepackage{graphicx}
\usepackage{epstopdf}
\usepackage{dcolumn}
\usepackage{bm}
\usepackage[sort&compress]{natbib}
\usepackage{graphicx, times}
\usepackage{color}

\begin{document}

\title{Circuit Quantum Electrodynamics with a Superconducting Quantum Point Contact}
\author{G. Romero$^1$, I. Lizuain$^{1,2}$, V. S. Shumeiko$^{3}$, E. Solano$^{1,4}$, and F. S. Bergeret$^{5}$}
\affiliation{$^{1}$ Departamento de Qu\'{i}mica F\'{i}sica, Universidad del Pa\'{i}s Vasco UPV/EHU, Apartado 644, 48080 Bilbao, Spain}
\affiliation{$^{2}$ Departamento de Matem\'atica Aplicada, Universidad del Pa\'{\i}s Vasco UPV/EHU, Apartado 644, 48080 Bilbao, Spain}
\affiliation{$^{3}$ Chalmers University of Technology, S-412 96 Got\"eborg, Sweden}
\affiliation{$^{4}$ IKERBASQUE, Basque Foundation for Science, Alameda Urquijo 36, 48011 Bilbao, Spain}
\affiliation{$^{5}$Centro de F\'{\i}sica de Materiales (CFM-MPC), Centro Mixto CSIC-UPV/EHU and Donostia International Physics Center (DIPC),
Manuel de Lardizbal 5, E-20018 San Sebasti\'an, Spain}

\date{\today}

\begin{abstract}
We consider a superconducting quantum point contact in a circuit quantum electrodynamics setup. We study three different configurations, attainable with current technology, where a quantum point contact is coupled galvanically to a coplanar waveguide resonator. Furthermore, we demonstrate that the strong and ultrastrong coupling regimes can be achieved with realistic parameters, allowing the coherent exchange between a superconducting quantum point contact and a quantized intracavity field.
\end{abstract}


\maketitle

{\it Introduction}.---A quantum point contact (QPC) is a constriction~\cite{Foxon88}, where electronic transport is supported by a small number of conducting channels, and the size of the constriction is smaller than the  inelastic mean free path.  In the normal state the resistance of the contact  is determined by the transmission of those channels. 
If the contact is made of a superconducting metal, the dissipationless current is carried by the Andreev bound states (ABS) localized at the junction region over the superconducting coherence length~\cite{Furusaki1991,Beenakker1991,Desposito2001,Zazunov2003,Zazunov2005,Shumeiko1993,Gorelik1995,Skoldberg2008,Bergeret2010,Koops1996,Goffman2000,Rocca2007,Zgirski2011,MChauvin2005,QLeMasne2009,ArxivCristian}. There is a pair of ABS for each conducting channel and their energies, $\pm E_A$ depends on the phase difference $\phi$ across the junction and the transmission coefficient $\tau$ of the conducting channel $E_A(\phi,\tau)=\Delta\sqrt{1-\tau\sin^2(\phi/2)}$, where $\Delta$ is the order parameter of the bulk superconducting electrodes. The energies of ABS are defined with respect to the Fermi level, and the current carried by each  state is  given by $I_{\pm}=\pm(2e/\hbar)\partial E_A/\partial\phi$. The total dc Josephson current through a QPC, consisting of a single conducting channel, is given by the contribution of each ABS weighted by the state occupation, $I=I_-n(-E_A)+I_+n(E_A)$, where $n(E)$ is the Fermi distribution function. Recent experiments on QPC have shown the correctness of describing the Josephson effect in terms of the ABS~\cite{Rocca2007,Zgirski2011,MChauvin2005,QLeMasne2009}. In addition, a direct spectroscopy of the ABS was done in a carbon nanotube attached to superconducting electrodes by means of tunneling spectroscopy~\cite{Pillet2010}. 

\begin{figure}[b]
  \includegraphics[width=0.4\textwidth]{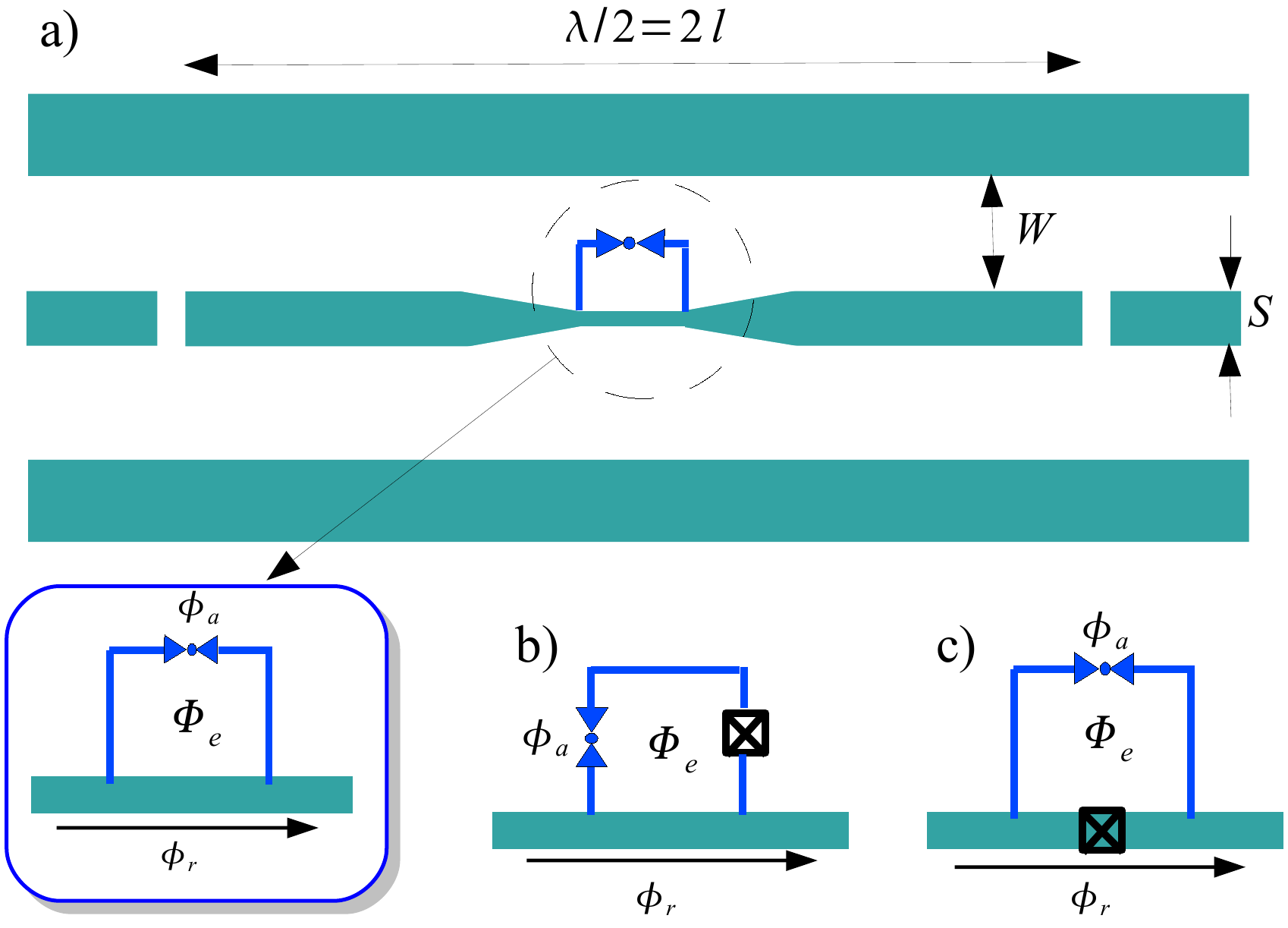}
  \caption{(Color online) (a) Superconducting loop containing a QPC  coupled galvanically to a CWR (see inset). (b) Atomic SQUID (aSQUID) with a QPC in series with a Josephson junction. (c)~aSQUID with the QPC in parallel with a Josephson junction.}
  \label{Fig1}
\end{figure} 

On the other hand, several theoretical  works have studied the ABS and their dynamics~\cite{Shumeiko1993,Gorelik1995,Desposito2001,Zazunov2003,Zazunov2005,Skoldberg2008,Bergeret2010}. In particular, it was recently proposed to use a QPC as a qubit~\cite{Zazunov2003}, introducing an effective Hamiltonian describing  the two-level system of  ABS. It has also been shown that in  the zero-temperature limit the dynamics of the QPC in a microwave field is well described by a two-level effective Hamiltonian \cite{Bergeret2010}. In this sense, it is predicted that the transition between the levels  induced by the microwave field  leads to a drastic suppression of the dc Josephson current.  However, though several attempts to detect such transitions have been carried out,  these have not been directly observed yet. In this context, it would be interesting to develop a theory involving a QPC coupled to a superconducting resonator in the context of state-of-the-art circuit quantum electrodynamics (QED)~\cite{Wallraff2004,Blais2004,Chiorescu2004,Hofheinz2009,Schoelkopf2011,Esteve2011,Cleland2011,Houck2011,Houck2012}. This may serve as an alternative physical scenario for probing the ABS dynamics, as well as a testbed for fundamental aspects in light-matter interaction. 

In this work, we propose a circuit QED setup using a superconducting QPC coupled galvanically~\cite{Abdumalikov2008} to a  coplanar waveguide resonator (CWR). Following previous works~\cite{Bourassa2009,Peropadre2010}, we develop a theoretical framework to study three experimental configurations shown in Fig.~\ref{Fig1}, and we present numerical simulations showing distinct cavity QED features as compared to the semiclassical case. Furthermore, we demonstrate that, for the configuration in Fig.~1c, one can reach the ultrastrong coupling (USC) regime of circuit QED~\cite{Bourassa2009,Peropadre2010,Ciuti05,Niemczyk10,Pol10}.  

{\it The Model}.---  Our generic setup consists of a superconducting loop containing a QPC coupled galvanically to a inhomogeneous resonator~\cite{Abdumalikov2008,Niemczyk10,Pol10} (see Fig.~\ref{Fig1}a).  We neglect the geometrical inductance of the loop, which we consider~\cite{QLeMasne2009} of the order of $20$~pH, with respect to the inductance of the QPC ($L_{QPC}\simeq 10$~nH)~\cite{Zgirski2011}.  
The system Hamiltonian reads 
\begin{equation}
H=H_{\rm QPC}(\phi_a)+H_{\rm CWR} \label{Ham}.
\end{equation}

We consider low temperatures $(T\!=\!0)$ and sufficiently small frequencies in order to suppress transitions between the ABS  and the continuum part of the spectrum. In this case, ss shown in Ref.~\cite{Bergeret2010},    one can describe the transport through the QPC by means of the  effective Hamiltonian derived by Zazunov {\it et al.}~\cite{Zazunov2003},
 \begin{equation}
H_{\rm QPC}(\phi_a) \! = \! \Delta e^{-i\sigma_x\sqrt{r}\phi_a/2} \! \left[\cos(\phi_a/2)\sigma_z \! + \! \sqrt{r} \sin(\phi_a/2)\sigma_y\right] \label{HQPCB} \! ,
\end{equation}
where  $r=1-\tau$ stands for the reflection coefficient and the Pauli matrices $\sigma_{x,y,z}$ are written  in the ballistic basis, defined by the eigenstates of the current for a perfectly transmitted conducting channel~\cite{QLeMasne2009}.  

The CWR is described by a sum of harmonic oscillators
\begin{equation} 
H_{\rm CWR}(\phi_n)=\sum_n \frac{1}{2}\bigg(\frac{2e}{\hbar}\bigg)^2\frac{\theta^2_n}{C_r}+\frac{1}{2}\bigg(\frac{\hbar}{2e}\bigg)^2C_r\omega^2_n\phi^2_n,
\end{equation} 
such that the total phase distribution on the CWR reads $\phi(x,t)=\sum_n u_n(x)\phi_n(t)$, where $u_n(x)$ is the spatial eigenmode. The phase variation along the segment of the resonator $\Delta x$, shared with the QPC loop, see Fig.~\ref{Fig1}a, reads $\phi_r=\sum_n\delta_n\phi_n$, where $\delta_n=u_n(x_0+\Delta x/2)-u_n(x_0-\Delta x/2)$ stands for the difference of the $n$th spatial eigenmode along $\Delta x$, and evaluated at the loop position $x_0$. The phase  $\phi_n$ is related to the annihilation and creation operators $\phi_n=(2e/\hbar)\sqrt{\hbar/(2\omega_nC_{r})}(a_n^\dagger+a_n)$, where $\omega_n$ is the corresponding eigenfrequency, and $C_r$ is the capacitance of the CWR; the conjugate momentum reads $\theta_n=i(\hbar/2e)\sqrt{\hbar\omega_n C_{r}/2}(a_n^\dagger-a_n)$. Typical resonator parameters are $C_r\sim 850~{\rm fF}$, the impedance $Z=\sqrt{L_r/C_r}\sim 50~\Omega$, where $L_r$ is the CWR inductance, and frequency $\omega_r/2\pi\sim 1-10~{\rm GHz}$ for the fundamental mode~\cite{Niemczyk10}.

In all cases displayed in Fig.~\ref{Fig1}, the description of the CWR in terms of eigenmodes is still valid if the inductance $L_{x}$ of the resonator segment $\Delta x$ is small as compared to the inductance of the QPC. Otherwise, the mode structure will depend strongly on the nonlinearities coming from the QPC current.  Under this approximation most of the current will flow through the resonator, and the QPC will act  as a small perturbation. For a CWR made of aluminum  with  width $S=50$~nm (at the constriction), thickness of the central electrode $t=50$~nm, length of the constriction  $\Delta x=5$~$\mu$m and  distance between the ground plane and the edge of the central line $W=4.95$~$\mu$m, we  estimate $L_{x} \simeq 8$~pH $\ll L_{\rm QPC}$, thus justifying the approximation. In order to simplify our analysis, we consider the simplest case where the QPC interacts with the lowest frequency eigenmode, $n=0$, supported by the resonator.

\begin{figure}
\includegraphics[width=0.45\textwidth]{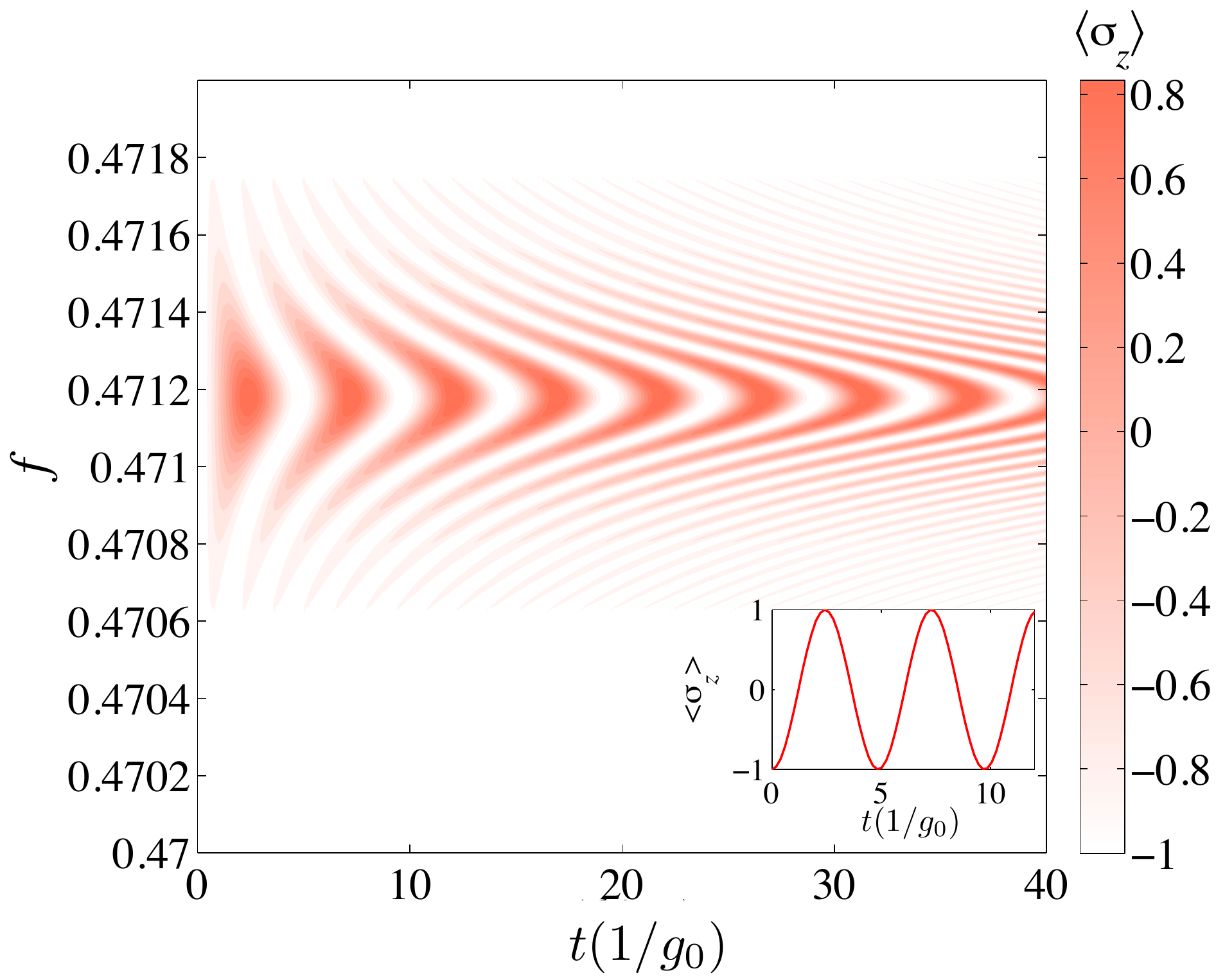}
\caption{(Color online) (a) Contour plot showing population inversion of Andreev levels for initial state $|\psi_0\rangle=|-\rangle|1\rangle_r$. The inset shows the mean value of $\sigma_z$ as a function of normalized time, $t(1/g_0)$, at the resonance condition $f_0=0.4712$.}
\label{Fig2}
\end{figure} 

The CWR and QPC are mutually coupled via the phase difference along the segment of the resonator $\Delta x$. Due to the flux quantization in superconducting loops, one can relate the external flux threading the loop, $\Phi$, with  the superconducting phase differences, $\sum_n\varphi_n=2\pi f+2\pi N$, where we have defined the frustration parameter $f=\Phi/\Phi_0$, being $\Phi_0=h/2e$ the flux quantum. In our treatment we will consider no trapped flux in the loop $(N=0)$ and we shall neglect tunneling effects coming from coherent phase slip~\cite{Tinkham2001,Harmans2005}. In particular, for the situation of Fig.~\ref{Fig1}a, the phase across the QPC ($\phi_a$) and the phase drop along the segment $\Delta x$ are related by $\phi_a=\phi_r + 2\pi f$. The superconducting phase difference $\phi_r$  is given by $\phi_r =\delta_0(2e/\hbar)(\hbar/2\omega_0C_r)^{1/2} (a + a^{\dag})$. We will also assume that $|\phi_r|\ll1$, a condition satisfied with current available values~\cite{Bourassa2009}. In this case, we can expand the Hamiltonian~(\ref{HQPCB}) up to first order on $\phi_r$. In addition, we rotate $H_{\rm QPC}$ in Eq. (\ref{HQPCB}) to the instantaneous  Andreev states basis\cite{Bergeret2010} by means of the transformation $H^{(A)}_{\rm QPC}(\phi_a)=UH^{(B)}_{\rm QPC}(\phi_a)U^\dagger$, where $U=e^{i\phi_a\sigma_y}e^{i\frac{\pi}{4}\sigma_z}$. Under this unitary transformation the whole Hamiltonian reads
\begin{equation}
H=\hbar \omega_0a^\dagger a+{E}_A(f,\tau)\sigma_z-\hbar g(a+a^\dagger)(A_z\sigma_z-A_x\sigma_x),\label{HQPCA}
\end{equation}
where $g=\frac{1}{2}\frac{\Delta^2}{\hbar}\tau\sin(\pi f)\delta_0(2e/\hbar)(\hbar/2\omega_0C_r)^{1/2}$ is the coupling strength, while $A_z$ and $A_x$ are the longitudinal and transversal coefficients, respectively, 

\begin{eqnarray}
A_z=\frac{\cos(\pi f)}{E_A(f,\tau)},~~
A_x=\frac{\sqrt{r}\sin(\pi f)}{E_A(f,\tau)}.
\end{eqnarray}
Notice that for a frustration  $f\sim0.5$, a large (small) contribution of the transversal (longitudinal) coupling is obtained. This is important since in a small vicinity of $f=0.5$ the QPC and the resonator can exchange excitations, so that ABS spectroscopy is possible. On the other hand, away from this working point the QPC-resonator interaction is dominated by the longitudinal coupling ($A_z$), where the field of the resonator will be displaced depending on the state of the QPC. These features show the versatility of our setup in order to engineer the QPC-resonator coupling, and will be the base for studying the underlying physics in three different setups. 

{\it QPC coupled to a coplanar waveguide resonator}.---We first analyze the configuration of Fig.~\ref{Fig1}a, i.e. a superconducting loop containing a QPC galvanically coupled  to the CWR.  The dynamics of this setup is described  by the Hamiltonian in Eq.~(\ref{HQPCA}).  This resembles the coupling of  a spin to a magnetic field with axial and longitudinal components and has the same  structure  like the  Hamiltonian describing  Cooper pair boxes or flux qubits~\cite{Wallraff2004,Blais2004,Chiorescu2004,Hofheinz2009,Abdumalikov2008,Bourassa2009,Peropadre2010,Niemczyk10,Pol10}.  It is worth mentioning that for parameters satisfying the conditions: $\omega_0 + 2E_A({\it f},\tau)/\hbar \gg \{gA_x,|\omega_0 - 2E_A({\it f},\tau)/\hbar|\}$ and $\omega_0 \gg gA_z$, the Hamiltonian~(\ref{HQPCA}) can be approximated by the 
the Jaynes-Cummings model~\cite{JaynesCummings} for which the rotating wave approximation holds.

We have studied numerically the dynamics of the Hamiltonian in Eq.~(\ref{HQPCA}) for an initial state having a single excitation in the resonator and the QPC in the ground state, i.e. $|\psi_0\rangle = \ket{-}\ket{1}_r$. Figure~\ref{Fig2} shows a contour plot of $\langle\sigma_z\rangle$ as a function of a normalized time, and the external flux applied to the loop close to the resonance condition, $\omega_0=2(\Delta/\hbar)(1-\tau\sin^2(\pi f))^{1/2}$, that leads to the working point
\begin{equation}
 f_0 = \frac{1}{\pi} \arcsin\Big[\Big( \frac{1}{\tau}\Big[1 -\Big(\frac{\hbar\omega_0}{2\Delta}\Big)^{2}\Big]\Big)^{1/2}\Big].
 \label{resonance}
 \end{equation}    
This figure reveals the population inversion of Andreev levels near  the working point $f_0$, as expected from the Jaynes-Cummings dynamics. This feature is more clearly shown in the inset of Fig.~\ref{Fig2}, where we  show the time evolution of  $\langle\sigma_z\rangle$ at the resonance  $f_0=0.4712$. In this simulation the CWR is described by a single mode with frequency $\omega_0/2\pi=10.52$~GHz, and a phase drop $|\phi_r| \sim 0.0013$ (see Ref.~\onlinecite{Bourassa2009}). For the QPC made of aluminum, we consider only one conducting  channel with high transmission, $\tau=0.994$,  and $\Delta/h\sim 44.256$~GHz (corresponding to a gap $\Delta\sim 0.183$~meV for Al)~\cite{QLeMasne2009}. These values lead to a ratio $g_0/\omega_0\Delta \sim 0.002$, where $g_0=\frac{1}{2}\frac{\Delta^2}{\hbar} \tau\delta_0(2e/\hbar)(\hbar/2\omega_0C_r)^{1/2}$. In principle, for these particular values the rotating-wave approximation can be safely used.  
 
\begin{figure}
\includegraphics[width=0.43\textwidth]{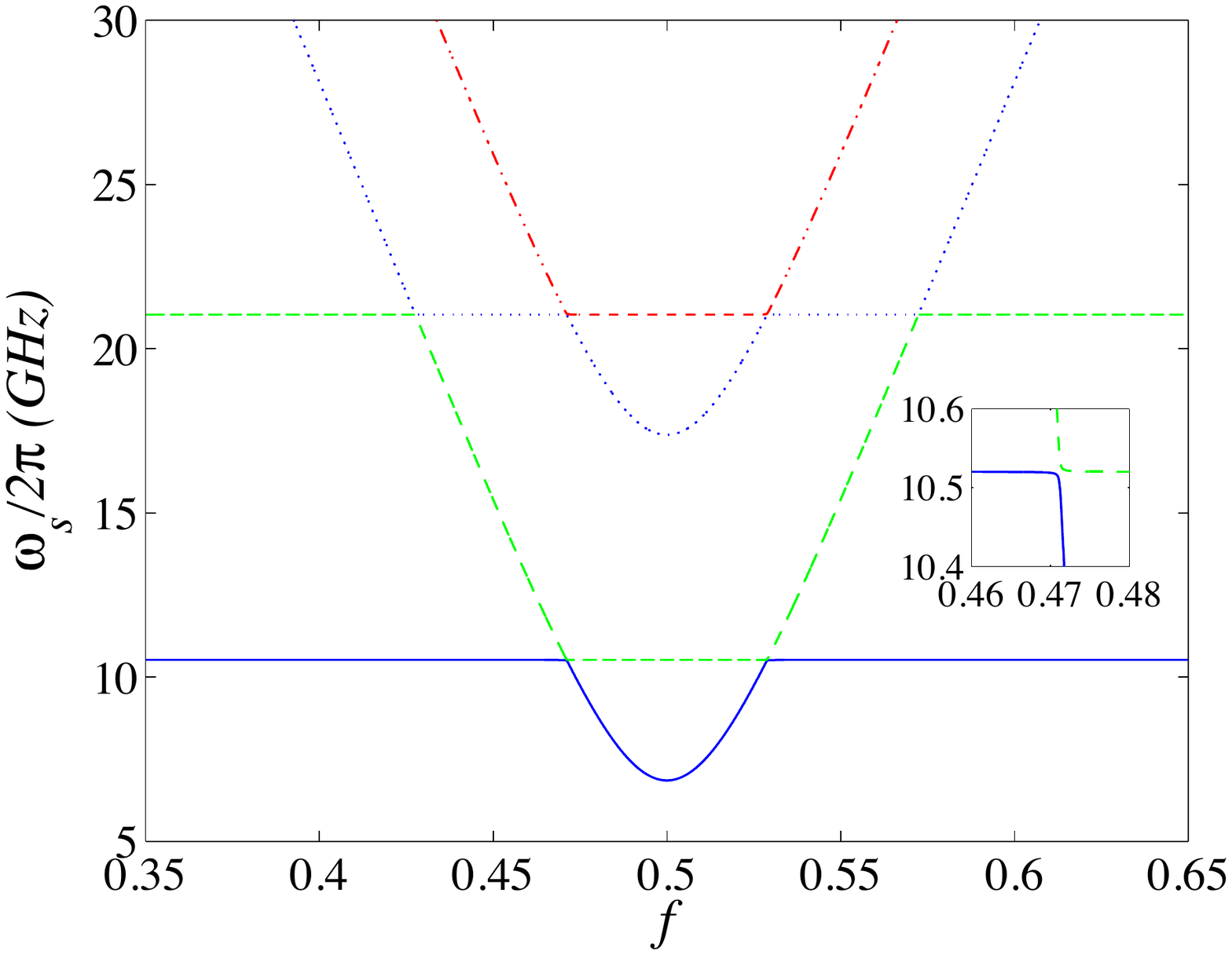}
\caption{(Color online) Energy spectrum for the configuration showed in Fig.~\ref{Fig1}a. The color codes stand for the energy differences---computed numerically--- solid-blue line (solid line) stands for $(E_1-E_0)/\hbar$, dashed-green line (dashed line) for $(E_2-E_0)/\hbar$, dotted-blue line (dotted line) for $(E_3-E_0)/\hbar$, and dot-dashed-red line (dot-dashed line) for $(E_4-E_0)/\hbar$. The inset shows the characteristic anticrossings representative of the Jaynes-Cummings model.} 
 \label{Fig3}
\end{figure}

It is worth mentioning that the vacuum Rabi oscillations shown in Fig.~\ref{Fig2} involves a time domain measurement of the Andreev levels. This can be done by, firstly, applying an external field via an on-chip flux line on the loop containing the QPC to far detune it from the resonator. Then, an amplitude shaped flux pulse $f_{\rm dc}$ is applied to tune the qubit into resonance with the cavity field for a variable time $\tau$. Then, probe the cavity frequency in a dispersive regime where the shift of the resonator frequency is proportional to the Andreev level population imbalance~\cite{QND}. In addition, the spectrum of the QPC-resonator system is shown in Fig.~\ref{Fig3}, that should be obtained via a cavity transmission measurement at frequency $\omega_s$~\cite{Niemczyk10}. The inset shows the avoided crossing between the single photon resonator state and the excited Andreev level as the QPC is brought into resonance at $f_0=0.4712$. The spectrum also shows avoided crossings coming from higher excitation states around $f = 0.4276$, yielding  a two-photon transition. A generalized expression for higher order  resonances is given by [{\it cf.} Eq. (\ref{resonance})],  $f_0=\frac{1}{\pi} \arcsin\Big[\Big( \frac{1}{\tau}\Big[1 -\Big(\frac{N\hbar\omega_0}{2\Delta}\Big)^{2}\Big]\Big)^{1/2}\Big]$ ($N=1,2,\dots$), which corresponds to possible  multi photon transitions in the the present setup.
  
{\it Atomic SQUID coupled to a coplanar waveguide resonator}.---For a proper characterization of the QPC and to control its phase, experiments have been carried out on an asymmetric SQUID loop consisting of a Josephson tunnel junction (JJ) in series with the QPC~\cite{Rocca2007,Zgirski2011,QLeMasne2009}. Here, we consider such an  atomic SQUID (aSQUID) galvanically coupled to the resonator as shown in Fig.~\ref{Fig1}b. In this configuration, the relation among the phases associated to the QPC ($\phi_a$), the  Josephson junction ($\phi_p$), and $\phi_r$ reads $\phi_a = \phi_r + \phi_p + 2\pi f$. The experiments considering this setup~\cite{Rocca2007,Zgirski2011,QLeMasne2009}, implement a large Josephson junction such that the Josephson energy is larger than the charging energy, $E_J \gg E_C$. In this case, the JJ-phase experiences small fluctuations and can be described by a harmonic oscillator characterized by annihilation and creation operators $b, b^{\dag}$, and plasma frequency $\omega_p = (2eI_C/\hbar C_J)^{1/2}$ with $C_J$ and $I_C$ the Josephson junction capacitance and critical current, respectively. Under this assumption we can expand the Hamiltonian, Eq.~(\ref{Ham}),  around a minimum.
Within this description the  Hamiltonian for the setup of Fig.~\ref{Fig1}b  reads 
\begin{eqnarray}
H& =& \hbar \omega_0a^{\dag}a+\hbar \omega_p b^{\dag}b + E_A(f,\tau)\sigma_z \nonumber \\ 
&-&\tau \frac{\Delta^2}{2}\sin(\pi f) (A_z\sigma_z-A_x	\sigma_x)(\phi_r+\phi_p)\nonumber \\
&-&\tau\frac{\Delta^2}{8E_A(f,\tau)}\bigg[(\cos^2(\pi f)-r\sin^2(\pi f))\sigma_z\nonumber\\
&-&2\sqrt{r}\sin(\pi f)\cos(\pi f)\sigma_x\bigg](\phi_r+\phi_p)^2,
\label{Hamiltonian_b}
\end{eqnarray}
where we have defined $\phi_p=(2e/\hbar)(\hbar/2C_J\omega_p)^{1/2}(b+b^{\dag})$.

\begin{figure}
\includegraphics[width=0.5\textwidth]{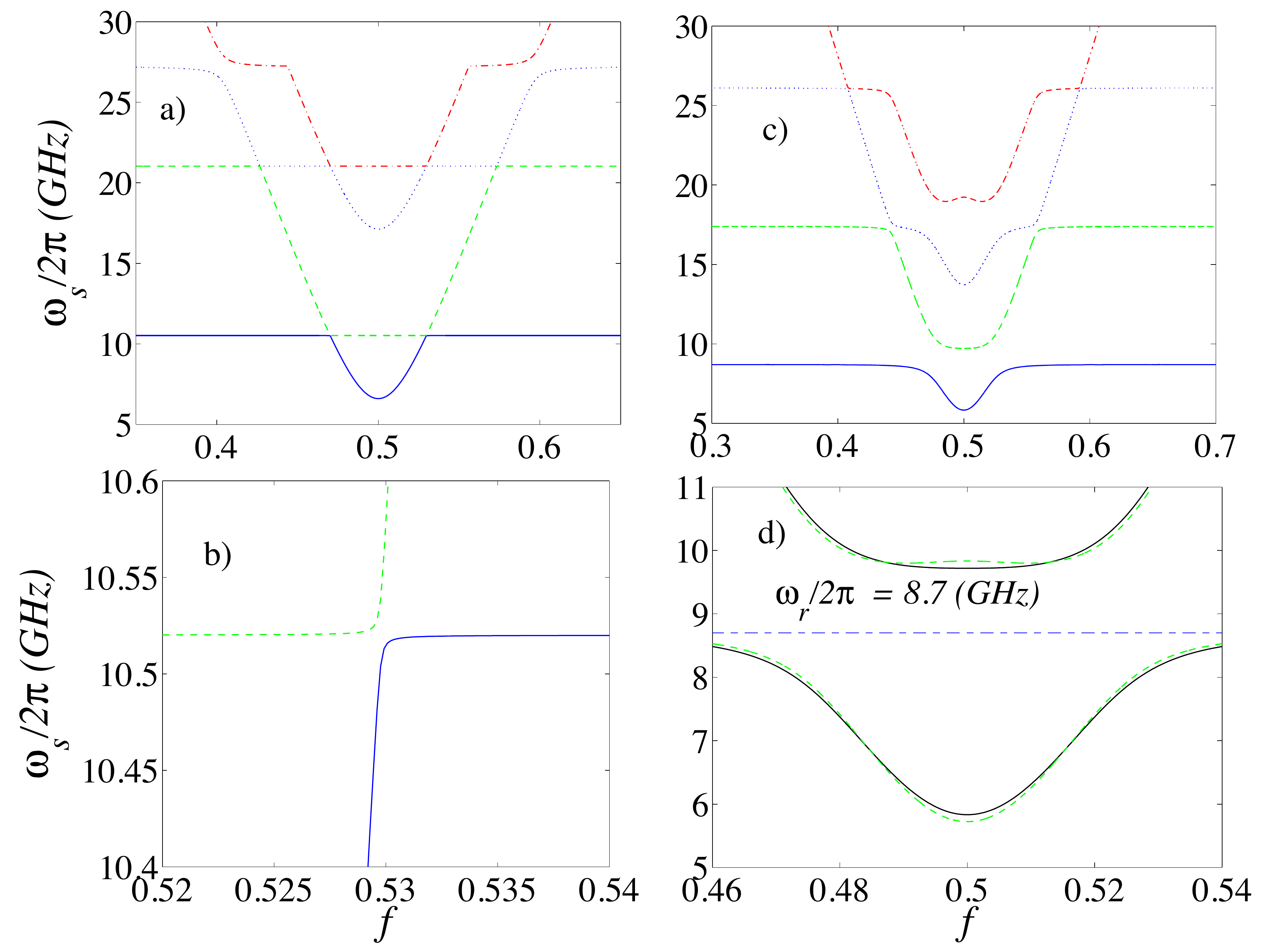}
\caption{(Color online) Energy spectrum for two cases of Fig.~\ref{Fig1}: (a) and (b) correspond to Fig. 1b, while (c) and (d) to Fig. 1c. The color codes represent the computed energy differences. For Fig.~\ref{Fig4}a,b,c the solid-blue line (solid line) stands for $(E_1-E_0)/\hbar$, dashed-green line (dashed line) for $(E_2-E_0)/\hbar$, dotted-blue line (dotted line) for $(E_3-E_0)/\hbar$, and dot-dashed-red line (dot-dashed line) for $(E_4-E_0)/\hbar$. The color codes in Fig.~\ref{Fig4}d are specified in the main text.} 
  \label{Fig4}
\end{figure}

In obtaining the Hamiltonian~(\ref{Hamiltonian_b}), we considered realistic values of $\phi_r$ and $\phi_p$ up to second order.  For instance, by taking a Josephson junction with critical current $I_C\sim 1 \mu$A, capacitance $C_J\sim100$ fF, and plasma frequency $\omega_p/2\pi\sim 27$~GHz~\cite{privatecomm}, one obtains $|\phi_p|\sim0.16$, while for the  the phase drop at the constriction we take  $|\phi_r| \sim 0.0013$. Notice that the dynamics results into a Stark shift of the Andreev levels depending on the number of excitations in the JJ and the CWR. Figures~\ref{Fig4}a and ~\ref{Fig4}b show the spectrum for the setup of Fig.~\ref{Fig1}b, calculated via a numerical diagonalization of Hamiltonian~(\ref{Hamiltonian_b}). It shows that the presence of the JJ induces an ac-Stark shift of the Andreev levels as depicted in Fig.~\ref{Fig4}b. We have estimated numerically a renormalized resonance condition,  $\bar{f}_0 \sim 0.47059$ where the QPC exchanges excitations with the resonator, as observed in the displayed anticrossings.

{\it Ultrastrong coupling regime}.--- Let us  consider the setup shown in  Fig.~\ref{Fig1}c, where the Josephson junction is now part of the resonator line. This setup is characterized with two main effects: (i) the renormalization of resonator eigenfrequencies, which are now given  by  the dispersion relation\cite{Bourassa2009,Wallquist06} $k_n = (2L_0/L_J)\big(1 - \omega^2_n/\omega^2_p\big)\cot(k_n l)$, where $L_J$ is the  junction inductance, $L_0$ the inductance per unit length of the resonator, and $l$ is half of cavity length. (ii) An increase of the coupling strength $g$  due to the local modification of the inductance of the single-mode resonator. For instance, if one takes $L_J\sim0.8$~nH, $C_J \sim 20$~fF, one obtains $\omega_p \sim 2\pi \times 40$~GHz and a resonator frequency $\omega_0 \sim 2\pi\times 8.7$~GHz. These values imply a phase drop $|\phi_r| = 0.0835$ such that the strength coupling of the QPC-resonator interaction  at the resonance phase in Eq.~(\ref{resonance}) reaches $g/\omega_0\Delta \sim 0.21$, well situated in the ultrastrong coupling  regime of light-matter interaction. This value is not restrictive and one we may reach ratios $g/\omega_0\Delta > 0.2$ for suitable parameters of the Josephson junction, and realize the regimes that have received a great theoretical and experimental attention in last years~\cite{Ciuti05,Bourassa2009,Niemczyk10,Pol10,Casanova10}.  Figures~\ref{Fig4}c and~\ref{Fig4}d show spectra for the USC coupling strength $g/\omega_0\Delta \sim 0.21$.  The solid black lines in  Fig.~\ref{Fig4}d  correspond to the spectrum calculated from Hamiltonian~(\ref{HQPCA}), while the dashed green lines represent the Jaynes-Cummings model.  The discrepancy between the two models is due to the  Bloch-Siegert shift introduced by the counterrotating terms in the full Hamiltonian~(\ref{HQPCA}). This shift was experimentally observed in a circuit QED setup with a flux quit~\cite{Pol10}, where a ratio $g/\omega_0\Delta=0.1$ has been reached. In this sense, we pave the way to study quantum optical properties associated to the QPC-CWR coupling in circuit QED.

Finally, note that dissipative mechanisms affecting the QPC coherence have already been studied~\cite{Desposito2001,Zazunov2005}. The theoretical estimated relaxation and dephasing rates are of the order of $10^8$ Hz. However this value can be substantially decreased by proper control of the  electromagnetic environment~\cite{privatecomm}.

{\it Conclusions.}---We have presented a general frame for studying the quantum dynamics of a superconducting point contact galvanically coupled to a single-mode resonator. We have analyzed three possible configurations that could be experimentally implemented and exhibit distinct cavity QED features as compared to semiclassical models. In particular, we have shown that the proposed QPC-CWR dynamics could reach the strong and ultrastrong coupling regimes of light-matter interaction, allowing the proposed setup to be considered as an alternative quantum device for circuit QED technology. 

We thank C. Urbina for discussions. We acknowledge funding from Spanish MICINN Juan de la Cierva, FIS2009-12773-C02-01, and FIS2011-28851-C02-02; Basque Government IT472-10 and IT- 366-07; UPV/EHU UFI 11/55; SOLID, CCQED, and PROMISCE European projects.

\end{document}